\documentclass[12pt]{article}

\usepackage{color}
\usepackage{amsmath}
\usepackage{graphicx}
\usepackage{amssymb}
\usepackage{amsthm}

\numberwithin{equation}{section}

\title{Spontaneous stochasticity of velocity \\in turbulence models}

\author{Alexei A. Mailybaev\footnote{Instituto 
Nacional de Matem\'atica Pura e Aplicada -- IMPA, 
Est. Dona Castorina 110, 22460-320 Rio de Janeiro, RJ, Brazil. 
E-mail: alexei@impa.br.} 
}

\date{}

\begin{document}

\maketitle

\begin{abstract}
We analyze the phenomenon of spontaneous stochasticity in fluid dynamics formulated as the nonuniqueness of solutions resulting from viscosity at infinitesimal  scales acting through intermediate on large scales of the flow. We study the finite-time onset of spontaneous stochasticity in a real version of the GOY shell model of turbulence. This model allows high-accuracy numerical simulations for a wide range of scales (up to ten orders of magnitude) and demonstrates non-chaotic dynamics, but leads to an infinite number of solutions in the vanishing viscosity limit after the blowup time. Thus, the spontaneous stochasticity phenomenon is clearly distinguished from the chaotic behavior in turbulent flows. We provide the numerical and theoretical description of the system dynamics at all stages. This includes the asymptotic analysis before and after the blowup leading to universal (periodic and quasi-periodic) renormalized solutions, followed by nonunique stationary states at large times. 
\end{abstract}


\section{Introduction}

One of important attributes of the developed hydrodynamic turbulence is the spontaneous stochasticity phenomenon. Spontaneous stochasticity is known for fields transported by turbulent flow~\cite{falkovich2001particles}, where non-Lipshitz inviscid velocity field leads to nonunique particle trajectories; see also \cite{eyink2015spontaneous} for a similar phenomenon for the backward in time Burgers equation. As the trajectories are unique at early times, when the flow is developing from large-scale smooth initial conditions, the finite-time onset of spontaneous stochasticity can be expected. Numerical analysis of this effect is substantially limited by the extreme multiscale nature of the problem.  
The possibility of the spontaneous stochasticity in incompressible flows is closely related to the existence of a finite-time blowup in the 3D Euler equations, which is the long-standing open problem of fluid dynamics and applied mathematics~\cite{gibbon2008three}. Though there exists a controversy on whether the blowup was observed in numerical studies, see e.g.~\cite{bustamante20083d,grafke2008numerical,hou2008blowup,agafontsev2015}, simulations show that the blowup can be triggered by a physical boundary~\cite{luo2014}. 

As it is common in theoretical studies, simplified (toy) models are useful for understanding complex phenomena in realistic systems. One of such popular models in hydrodynamic turbulence is the Gledzer-Ohkitani-Yamada (GOY) shell model~\cite{gledzer1973system,ohkitani1989temporal}, modified later into the Sabra model~\cite{l1998improved}. This model describes characteristic velocity variations $u_n$ at different scales considered in a geometric progression with the corresponding wave numbers $k_n = k_02^n$. The dynamics is governed by an infinite (truncated for numerical purposes) system of ordinary differential equations, which possesses several properties of the Navier-Stokes equations such as scaling symmetries and conserved quantities, and provides the chaotic behavior with the energy cascade between large and small scales. Though the GOY and Sabra models are highly convenient for numerical analysis (allowing scale resolution over ten orders of magnitude) and a large amount of knowledge on their behavior is collected~\cite{biferale2003shell}, their dynamics is still far from being completely understood theoretically.

The blowup problem for the inviscid shell models was solved~\cite{dombre1998intermittency,mailybaev2012c}, showing the existence of a finite-time blowup with a universal asymptotic form. A clear evidence is also given about the role of blowup in the turbulent energy transport to small scales~\cite{gilson1997towards,mailybaev2012computation,mailybaev2013blowup}. 
In this paper, our goal is the numerical and theoretical description of the dynamics after the blowup in the vanishing viscosity limit. For this purpose, we use the real version of the (originally complex) GOY model. This model demonstrates the deterministic behavior before the blowup time, followed by spontaneous stochasticity after the blowup. Unlike its complex version, this shell model is not chaotic, thus, clearly distinguishing the spontaneous stochasticity from  chaos in dynamical systems. From this perspective, the proposed model can be considered as an the intermediate step between the dyadic shell models~\cite{desnyansky1974evolution,cheskidov2007inviscid}, whose behavior is deterministic and resembles the Burgers equation~\cite{mailybaev2014continuous}, and the chaotic complex GOY or Sabra models, whose inviscid limit is not yet understood. 

The contribution of this work is the description of the inviscid limit leading to an infinite number of solutions, triggered by vanishing viscosity through the whole range of scales. The specially nontrivial part of this analysis is the theory explaining the onset of spontaneous stochasticity, developing after the blowup. Using the renormalization technique, we show the existence of an asymptotic solution in the form of a universal quasi-periodic pulse moving from infinitely small to large scales in finite time. The obtained results show that the nonuniqueness in inviscid flows with blowup can be described by considering viscosity as a small random variable, while keeping deterministic boundary and initial conditions. This leads to unique and physically relevant inviscid solutions in the form of probability measures, providing a new insight on the multiscale behavior of developed turbulence.

The paper is organized as follows. We start with the model description in Section~\ref{sec:Model}. Then we describe viscous stationary solutions in Section~\ref{sec3} and their inviscid limit in~Section~\ref{sec4}. These results are used in Section~\ref{sec5} for defining the time-dependent inviscid solutions, which appear to be nonunique and depend on the specific form of  inviscid limit. Section~\ref{sec6} describes the blowup structure. Section~\ref{sec7} develops the renormalization approach providing the universal asymptotic quasi-periodic solution after the blowup governing the onset of spontaneous stochasticity. We end with some conclusions.

\section{Model and inviscid stationary solutions}
\label{sec:Model}

According to the Kolmogorov--Obukhov theory~\cite{kolmogorov1941local,obukhov1941spectral}, chaotic dynamics 
of the incompressible 3D Navier-Stokes equations
\begin{equation}
\frac{\partial \mathbf{u}}{\partial t}
+\mathbf{u}\cdot \nabla \mathbf{u} = -
\nabla P +\nu\Delta\mathbf{u},\quad
\nabla\cdot\mathbf{u} = 0,
\label{eqNS}
\end{equation}
for large Reynolds numbers can be described using the concept of isotropic homogeneous turbulence. In this theory, the energy is transported from large (forced) to small (viscous) scales through a large inertial interval of intermediate scales $\ell$, where the 
velocity fluctuations obey the power-law  
$\langle|\delta u|\rangle\propto \varepsilon^{1/3} \ell^{1/3}$ with  
the mean energy flux $\varepsilon$ from large to small scales. Such description, based on the dimensional analysis is known to be incomplete due to the anomalous corrections (not yet explained theoretically) for the exponents of this and higher velocity moments, see e.g.~\cite{landau2013fluid,frisch1999turbulence}. Because of the extremely large number of degrees of freedom for realistic Reynolds numbers, this problem is very difficult for the numerical analysis.

In this paper, we consider the shell model of turbulence given by an infinite system of differential equations
\begin{equation}
\frac{d u_n}{d t}
=
\left(\frac{1}{2}k_{n-1}u_{n-1}u_{n-2}
+\frac{1}{2}k_nu_{n+1}u_{n-1}
-k_{n+1}u_{n+2}u_{n+1}\right)
-\nu k_n^2 u_n
\label{eq1}
\end{equation}
for $n = 1,2,\ldots$, with constant boundary conditions
\begin{equation}
u_0 = const,\quad u_{-1} = const. 
\label{eq1b}
\end{equation}
In this system, the real quantities $u_n$ model qualitatively the Fourier components of a velocity field $\mathbf{u}(\mathbf{k},t)$ corresponding to wavenumbers $|\mathbf{k}| = k_n$ (one can also consider $u_n$ as a characteristic velocity fluctuation at spatial distance $\ell \sim 2\pi/k_n$), the wavenumbers $k_n = k_02^n$ are chosen in the form of geometric progression, and $\nu$ is the viscosity parameter. System (\ref{eq1}) represents the Gledzer-Ohkitani-Yamada shell model~\cite{gledzer1973system,ohkitani1989temporal} taken for purely imaginary shell velocities $-iu_n$; similarly, it can be deduced from the Sabra model of turbulence~\cite{l1998improved}. For simplicity, we set $k_0 = 1$ and consider the boundary conditions $u_{-1}^2+u_0^2 \sim 1$ by an order of magnitude.
The model (\ref{eq1}) is considered as a ``toy model'' for 3D Navier-Stokes turbulence in Eq.~(\ref{eqNS}), as it possesses the same scaling symmetry and shares some of the motion invariants~\cite{biferale2003shell}. In particular, the quadratic terms in (\ref{eq1}) mimic the convective nonlinearity and the pressure term in Eq.~(\ref{eqNS}). The clear advantage of the shell model is the drastic reduction in the number of degrees of freedom, while keeping the infinite-dimensional nature of the flow. 

Let us define the ``energy'' as $E = \sum u_n^2$ and the ``enstrophy'' as $\Omega = \sum k_n^2u_n^2$. 
Then the energy balance equation takes the form similar to the Navier-Stokes equations (see e.g.~\cite{frisch1995turbulence}) as
\begin{equation}
\frac{dE}{dt} = \Pi_0-2\nu\Omega,
\label{eq1c}
\end{equation}
where 
\begin{equation}
\Pi_n = k_{n}u_{n-1}u_{n}u_{n+1}+2k_{n+1}u_{n}u_{n+1}u_{n+2}
\label{eq1d}
\end{equation}
is the energy flux between the shells with wavenumbers $k_n$ and $k_{n+1}$. Thus, the term $\Pi_0$ in Eq.~(\ref{eq1c}) represents the energy flux into the system generated by boundary conditions (\ref{eq1b}), while the term $2\nu\Omega$ describes the viscous energy dissipation. A similar balance law holds for the helicity invariant introduced as $H = \sum (-1)^nk_nu_n^2$, see e.g.~\cite{biferale2003shell}. 

The following three transformations 
\begin{equation}
u_n \mapsto 2u_{n+1}, \quad \nu \mapsto 2^2\nu;  
\label{eqS1}
\end{equation}
\begin{equation}
u_n \mapsto cu_n, \quad \nu \mapsto c\nu,\quad t\mapsto t/c;  
\label{eqS2}
\end{equation}
\begin{equation}
u_n \mapsto \sigma_nu_n, \quad \sigma_n = \pm 1, \quad 
\sigma_n\sigma_{n+1}\sigma_{n+2} = 1, 
\label{eqS3}
\end{equation}
are the symmetries of system (\ref{eq1}). Also, the system is invariant under time translations $t \mapsto t+t_0$. Note that the shift $n \mapsto n+1$ implies the scaling $k_n \mapsto k_{n+1} = 2k_n$ in Fourier space. Therefore, Eqs.~(\ref{eqS1}) and (\ref{eqS2}) are related to the scaling of space-time. The signs $\sigma_n$ in Eq.~(\ref{eqS3})  are periodic with $\sigma_n = \sigma_{n+3}$ and can be associated with phase factors in the Fourier transform induced by physical space translations, see e.g.~\cite{biferale2003shell}.

Stationary solutions of the inviscid model ($\nu = 0$) were described in~\cite{biferale1995transition}. The equilibrium 
conditions for Eqs.~(\ref{eq1}), after dividing by $k_{n+1}$, take the form
\begin{equation}
u_{n+2}u_{n+1}
= \frac{1}{4}u_{n+1}u_{n-1}
+\frac{1}{8}u_{n-1}u_{n-2},
\quad n = 1,2,\ldots.
\label{eq4}
\end{equation}
If $u_n \ne 0$ for all $n \ge 1$, then Eqs.~(\ref{eq4}) define recursively the values of all shell speeds $u_n$, $n \ge 3$, 
for given $u_{-1},u_0,u_1,u_2$. Equations~(\ref{eq4}) after multiplication by $2u_{n-1}^{-1}u_{n+1}^{-1}$ yield 
the recurrent relation for velocity ratios 
\begin{equation}
c_{n+2} = \frac{1}{2}+\frac{1}{2c_{n+1}},\quad
c_n = \frac{2u_{n}}{u_{n-3}},
\label{eq5}
\end{equation}
which possesses a single attracting fixed-point $c_n = 1$ as $n \to \infty$. 
This implies that $u_{3n+i}$ is asymptotically proportional to $2^{-n}$ for each $i = 1,2,3$.
Thus, with an increase of $n$, the stationary solutions have the period-3 asymptotic form
\begin{equation}
u_{3n+i} \to a_ik_{3n+i}^{-1/3},\quad 
i = 1,2,3,
\label{eq6}
\end{equation}
with constant real quantities $a_1,a_2,a_3$.

Note that the relation (\ref{eq6}) implies the power law
\begin{equation}
u_n \sim k_n^{-1/3},
\label{eq6b}
\end{equation}
which
can be associated with the Kolmogorov scaling law of developed turbulence mentioned above, as well as with the shock wave solutions in continuous models~\cite{mailybaev2014continuous}. 

\section{Viscous stationary solutions}
\label{sec3}

\begin{figure}
\centering
\includegraphics[width = 0.58\textwidth]{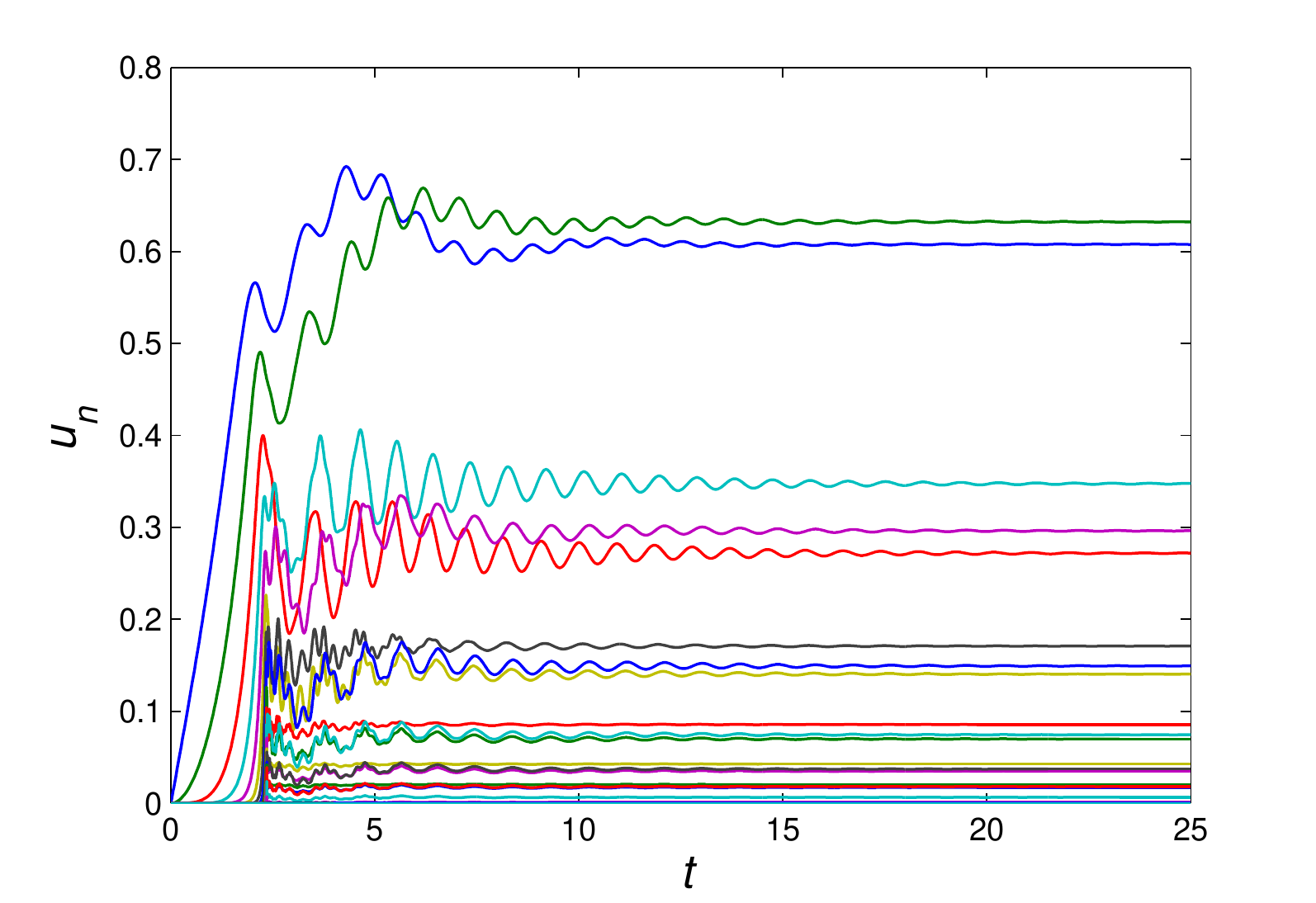}
\caption{Solution for the shell model for zero initial conditions, boundary values $u_{-1} = u_0 = 0.7$ and viscosity $\nu = 2^{-4N} \approx 6\times 10^{-8}$ with $N = 6$. Shown are the values of shell speeds $u_n(t)$ for $n = 1,2,\ldots$ converging to their stationary values; the speeds $u_n$ decrease (nonmonotonously) with increasing $n$.}
\label{fig1}
\end{figure}

For a positive viscosity, $\nu > 0$, and constant boundary conditions (\ref{eq1b}), numerical simulations show that the shell model 
(\ref{eq1}) possesses a fixed-point attractor, Fig.~\ref{fig1}. 
In this case, the equilibrium conditions become
\begin{equation}
u_{n+2}u_{n+1} = 
\frac{1}{4}u_{n+1}u_{n-1}
+\frac{1}{8}u_{n-1}u_{n-2}
-\frac{1}{2}\nu k_nu_n,\quad
n = 1,2,\ldots,
\label{eq7}
\end{equation}
which differ from Eq.~(\ref{eq4}) by an additional viscous term.
For nonzero shell speeds, Eq.~(\ref{eq7}) determines the equilibrium state recursively if 
the four initial speeds $u_{-1}$, $u_0$, $u_1$, $u_2$ are given. The speeds $u_{-1}$, $u_0$ are determined by the 
boundary conditions, while the speeds $u_1$, $u_2$ depend on the viscosity, 
as we will see below. 

\begin{figure}
\centering
\includegraphics[width = 0.55\textwidth]{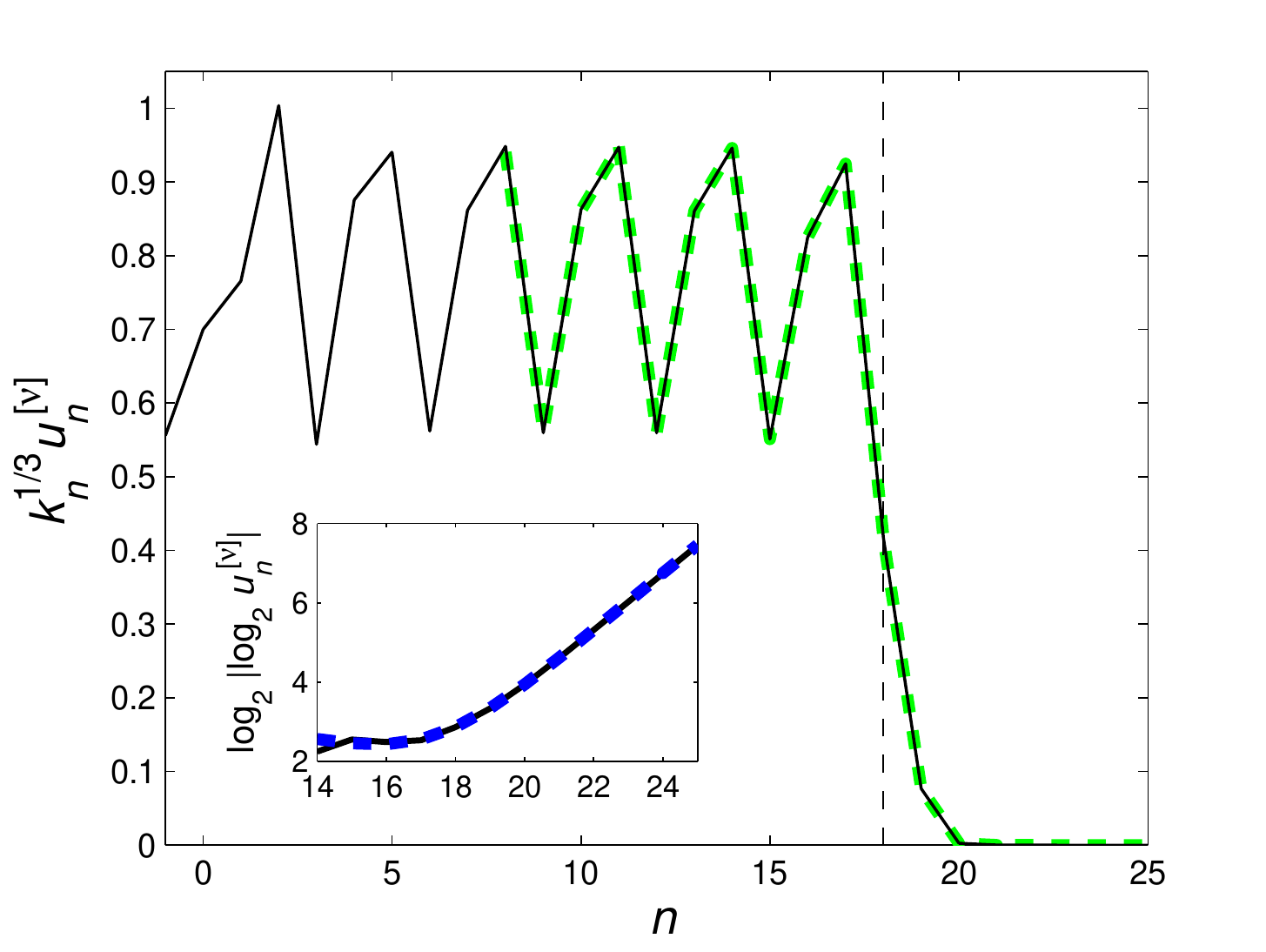}
\caption{Scaled shell speeds $k_n^{1/3}u_n^{[\nu]}$ of the stationary attractor found numerically for the solution $u_n(t)$ in Fig.~\ref{fig1}. The viscous range position $n_K = 18$ is shown by the vertical dashed line. The stationary solution is periodic in the inertial range $1 \ll n \ll n_K$. The dashed bold green line shows the viscous asymptotic  solution $k_{n-3N}^{1/3}V_{n-3N}(\chi)$ from Eq.~(\ref{eq12}), which determines the period-3 state (\ref{eq12b}) in the inertial range. The inset compares numerical results (solid black line) with the asymptotic expression (\ref{eq10}) (dashed blue line) for $b = -0.0011$; plotted are the values of $\log_2|\log_2u_n^{[\nu]}|$ for $n = 14,\ldots,25$. }
\label{fig2}
\end{figure}

For small viscosity $\nu$, the viscous term in Eq.~(\ref{eq7}) is small and the solution is 
determined approximately by the inviscid model (\ref{eq4}) leading to the properties (\ref{eq6}). 
However, for any finite $\nu > 0$, the viscous term becomes important for large shell numbers (small scales), 
when $\nu k_n \sim u_n$. Using this estimate with the scaling law (\ref{eq6b}), 
we introduce the Kolmogorov wavenumber as $k_K \sim \nu^{-3/4}$. The corresponding shell number 
\begin{equation}
n_K = \log_2k_K = -\frac{3}{4}\log_2\nu 
\label{eq7b}
\end{equation}
separates the region $n \ll n_K$ of inviscid dynamics from the viscous 
range $n \gtrsim n_K$, Fig.~\ref{fig2}. The inviscid region contains the forcing (boundary) range at $n \sim 1$ and the inertial interval $1 \ll n \ll n_K$. 

In the viscous range, one expects fast decay of velocities $u_n$ with increasing shell number $n$, due to the large factor $k_n^2$ of the viscous term in Eq.~(\ref{eq1}). In this case, the last two terms in Eq.~(\ref{eq7}) are dominant, which yields
\begin{equation}
u_{n-1}u_{n-2} \approx 4\nu k_nu_n.
\label{eq8}
\end{equation}
Writing this expression as a linear equation 
\begin{equation}
\log_2|u_n| \approx 
\log_2|u_{n-1}|+\log_2|u_{n-2}|-2-n-\log_2 \nu,
\label{eq9}
\end{equation}
one finds the general solution in the form
\begin{equation}
\log_2|u_n| \approx  
b\sigma^n+\tilde{b}\tilde{\sigma}^n+5+n+\log_2\nu
\label{eq9b}
\end{equation}
with $\sigma = (1+\sqrt{5})/2$, $\tilde\sigma = (1-\sqrt{5})/2$ and arbitrary real coefficients $b$ and $\tilde{b}$. Since $\sigma > 1$ and $|\tilde\sigma| < 1$, we find the asymptotic behavior as
\begin{equation}
|u_n| \approx 
32\nu k_n2^{b\sigma^n},\quad n \gg n_K,
\label{eq10}
\end{equation}
where the coefficient $b$ must be negative in order to have $|u_n| \to 0$ for large $n$ (similar analysis was carried out in~\cite{dombre1998intermittency} for a different shell model). 
Both asymptotic expressions (\ref{eq6}) for $n \ll n_K$ and (\ref{eq10}) for $n \gg n_K$ are confirmed numerically, see Fig.~\ref{fig2}.

\section{Inviscid limit for stationary solutions}
\label{sec4}

Let us denote by $u_n^{[\nu]}$ the stationary solution corresponding to the viscosity $\nu$, with fixed boundary conditions. 
For understanding the limit of vanishing viscosity, $\nu \to +0$,
we first study the asymptotic behavior in the viscous range, $n \gtrsim n_K$, 
given by Eq.~(\ref{eq7b}). 

\begin{figure}
\centering
\includegraphics[width = 0.55\textwidth]{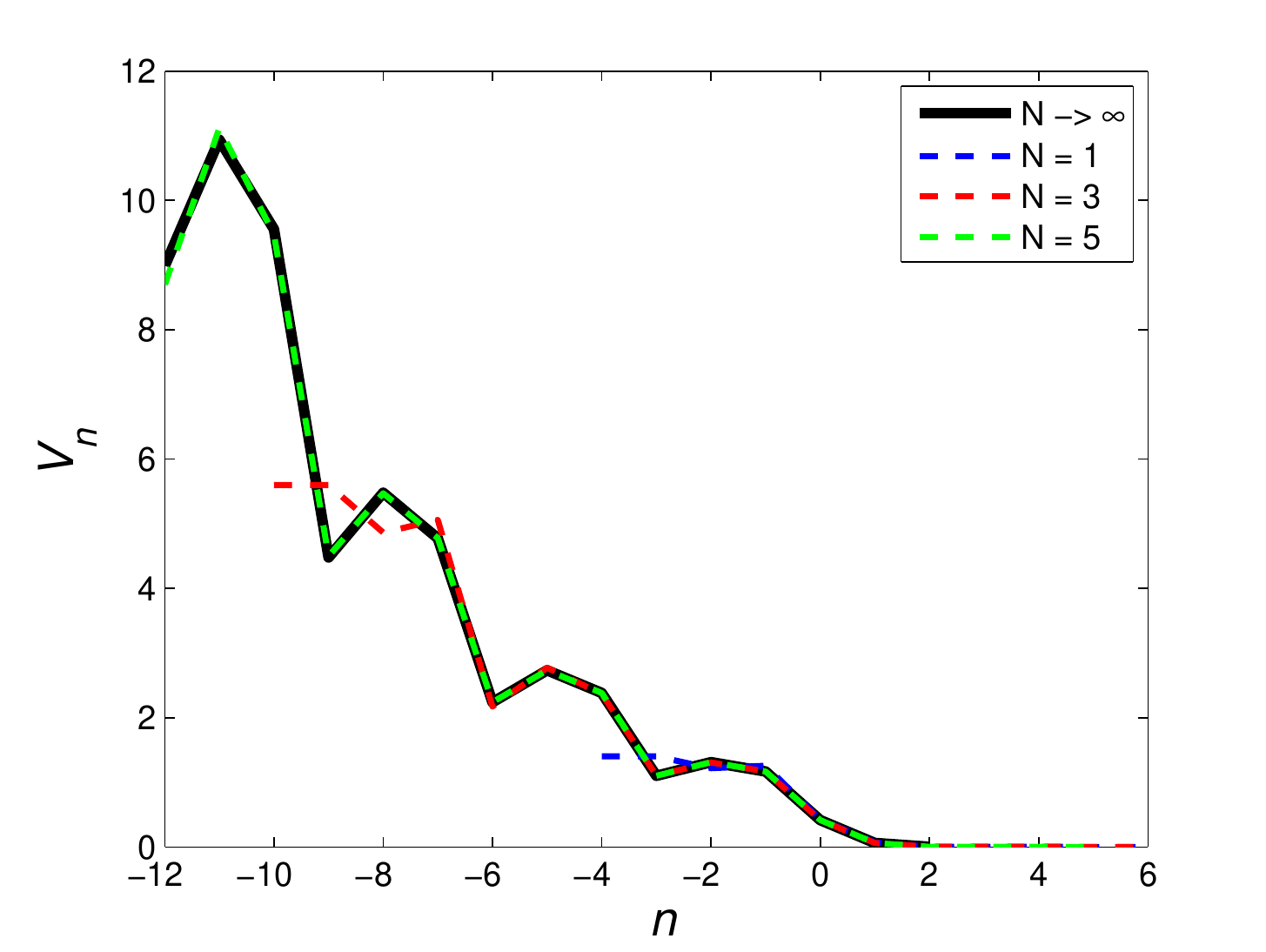}
\caption{Expression $2^Nu_{n+3N}^{[\nu_N]} \to V_n$ of the limit (\ref{eq12}) shown for $N = 1,3,5$ and $N \to \infty$ based on numerical simulations with the conditions of Fig.~\ref{fig1}.}
\label{fig3}
\end{figure}

For small viscosity, stationary solutions satisfy the period-3 asymptotic relation (\ref{eq6}) in the inertial range $1 \ll n \ll n_K$, where the coefficients $a_1,a_2,a_3$ depend on $\nu$. Expression (\ref{eq6}) is invariant under the scaling transformation 
\begin{equation}
u_n \mapsto 2^Nu_{n+3N},
\label{eq10b}
\end{equation}
with the shift by $3N$ shell numbers.
The viscous range $n \gtrsim n_K$ given by Eq.~(\ref{eq7b}) can be shifted back to its original position, 
$n_K \mapsto n_K-3N$, by changing the viscosity as
\begin{equation}
\nu \mapsto 2^{4N}\nu.
\label{eq10c}
\end{equation}
Thus, one can expect that the combination of transformations (\ref{eq10b}) and (\ref{eq10c}) leaves the stationary solution approximately unchanged, i.e.,
\begin{equation}
u_n^{[\nu]} \approx 2^Nu_{n+3N}^{[\nu_N]}, \quad 
\nu = 2^{4N}\nu_N.
\label{eq11}
\end{equation}
This expression is valid for all $n$ except for the boundary layer near $n = 1$, which is not shift-invariant. Since relation (\ref{eq11}) should be satisfied with higher accuracy for a smaller viscosity, we consider the limit
\begin{equation}
V_n(\chi) = \lim_{N \to \infty} 2^Nu_{n+3N}^{[\nu_N]}, \quad 
\nu_N = 2^{-4(\chi+N)}, \quad n,N \in \mathbb{Z},
\label{eq12}
\end{equation}
where $n$ is an arbitrary (positive or negative) integer and $\nu_N \to 0$ as $N \to \infty$ for a fixed parameter $\chi \in \mathbb{R}$. Existence of this limit is confirmed numerically in Fig.~\ref{fig3}. 
In Fig.~\ref{fig2} the solution $k_n^{1/3}u_n^{[\nu_N]}$ (solid line) is presented together with its viscous range asymptotic form $k_{n-3N}^{1/3}V_{n-3N}(\chi)$  (dashed green line) given by Eq.~(\ref{eq12}).   

\begin{figure}
\centering
\includegraphics[width = 1\textwidth]{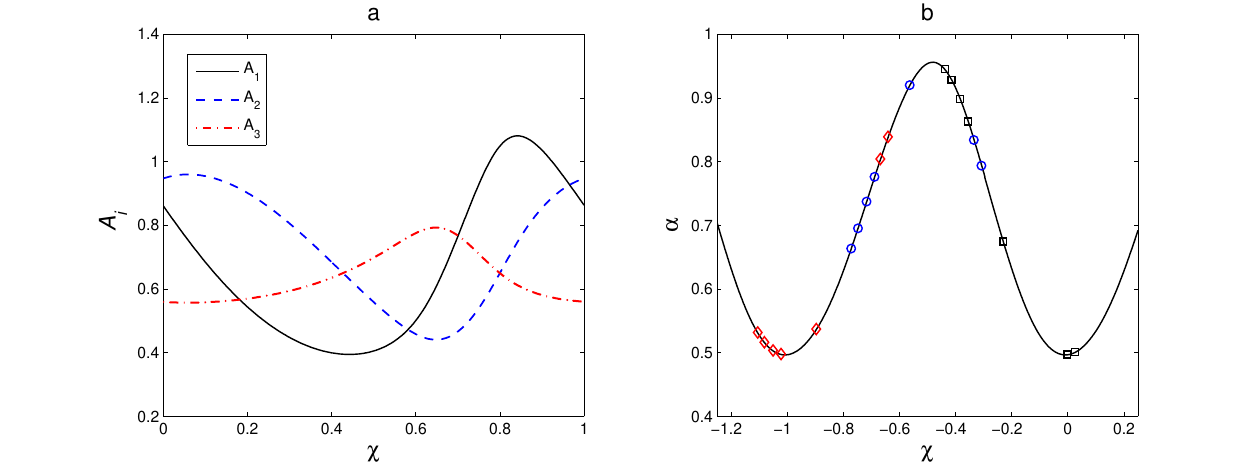}
\caption{a) Functions $A_i(\chi)$ determining the period-3 stationary state in the inertial interval depending on the parameter $\chi$ for boundary conditions of Fig.~\ref{fig1}. b) The function $\alpha(\chi)$ determining the universal period-3 stationary state (\ref{eq18}) in the inertial interval. Squares ($i = 1$), circles ($i = 2$) and diamonds ($i = 3$)  represent the values of 
$\alpha = D^{-1/3}k_{3n+i}^{1/3}|u_{3n+i}|$ for $n = 4$ versus $\chi-\frac{1}{3}+\frac{\log_2D}{12}$ obtained in numerical tests with random boundary conditions and random $|\chi| < 1/2$.}
\label{fig4}
\end{figure}

Combining Eqs.~(\ref{eqS1}) and (\ref{eqS2}) with $c = 2^{-2/3}$, we obtain the symmetry transformation
\begin{equation}
u_n \mapsto 2^{1/3}u_{n+1},\quad
\nu \mapsto 2^{4/3}\nu,\quad
t \mapsto 2^{2/3}t,
\label{eqS5}
\end{equation}
which leaves Eq.~(\ref{eq1}) invariant. Hence, 
expression of the limit in Eq.~(\ref{eq12}) represents the symmetry transformation (\ref{eqS5}) repeated $3N$ times for stationary solutions. In particular, this implies that  
$V_n(\chi)$ is a stationary solution for the viscous system with $\nu = 2^{-4\chi}$. 
The inertial range of $V_n(\chi)$ with the period-3 behavior (\ref{eq6}) extends to large negative $n$ (see Figs.~\ref{fig2} and \ref{fig3}) leading to the boundary condition
\begin{equation}
V_{3n+i}(\chi) \to A_i(\chi)k_{3n+i}^{-1/3}
\quad \textrm{as} \quad
n \to -\infty,\quad i = 1,2,3.
\label{eq12b}
\end{equation}
The coefficients $A_i$ depend on $\chi$ (see Fig.~\ref{fig4}a) and can be written using Eqs.~(\ref{eq12}) and (\ref{eq12b}) as
\begin{equation}
\begin{array}{c}
\displaystyle
A_i(\chi) = \lim_{n\to -\infty}
\lim_{N\to\infty}
k_m^{1/3}u_m^{[\nu_N]},\\[10pt]
\displaystyle
m = 3(n+N)+i,\quad
\nu_N = 2^{-4(\chi+N)}, \quad
i = 1,2,3.
\end{array}
\label{eq12d}
\end{equation}
The order of the limits in Eq.~(\ref{eq12d}) is important. It describes the period-3 stationary solution in the inertial interval $1 \ll m \ll n_K$, 
where $n_K = 3(\chi+N) \to \infty$ with vanishing viscosity $\nu_N \to 0$. By inspection of Eq.~(\ref{eq12d}), one can check that the relations
\begin{equation}
A_i(\chi+k) = A_i(\chi),\quad
i = 1,2,3, \quad k \in \mathbb{Z},
\label{eq12f}
\end{equation}
are valid for any integer $k$. 

The obtained results show that the period-3 asymptotic form (\ref{eq6}) depends on the parameter $\chi$ of the inviscid limit. This leads to the important conclusion that the inviscid limit $\lim_{\nu \to +0}u_n^{[\nu]}$ for stationary solutions does not exist in the usual sense, while one can define an infinite number of inviscid stationary solutions 
\begin{equation}
U_n(\chi) = \lim_{N \to \infty}u_n^{[\nu_N]},\quad
\nu_N = 2^{-4(\chi+N)},
\label{eq16}
\end{equation}
obtained for specific viscosity subsequences $\nu_N \to 0$ depending on the parameter $\chi$. These solutions satisfy the condition
\begin{equation}
U_{3n+i}(\chi) \to A_i(\chi)k_{3n+i}^{-1/3},\quad 
i = 1,2,3,\quad
n \to \infty,
\label{eq15}
\end{equation}
where the coefficients $A_i(\chi)$ are determined in Eq.~(\ref{eq12d}) by the viscous mechanism for specific boundary conditions. Since the values $\chi$ and $\chi+k$ with any integer $k$ yield the same inviscid solutions, the parameter $\chi$ should be considered modulo 1. 

Substituting expression (\ref{eq15}) into Eq.~(\ref{eq1d}), we find 
\begin{equation}
D(\chi) = \lim_{n \to \infty} \Pi_n 
= 3A_1(\chi)A_2(\chi)A_3(\chi)
\label{eq12c}
\end{equation}
for the energy flux in the inertial interval. This value is equal to the injected energy flux $\Pi_0$ because the viscous dissipation acts only at infinitesimal viscous scales (infinite shell numbers $n$). Since $dE/dt = 0$ for the stationary solution in Eq.~(\ref{eq1c}), we find that the value $D(\chi)$ represents the total energy dissipation rate of the inviscid solution $U_n(\chi)$. This phenomenon of finite dissipation in the inviscid limit is known as the dissipation anomaly~\cite{eyink2006onsager}.

We described the vanishing viscosity limit of stationary solutions under fixed boundary conditions, which were taken in the numerical simulations as $u_{-1} = u_0 = 0.7$. Using the symmetry (\ref{eqS2}) with an arbitrary coefficient $c > 0$, we see that $\tilde{U}_n = cU_n(\chi)$ is the inviscid solution obtained in the vanishing viscosity limit $\tilde{\nu}_N = c\nu_N = c2^{-4(\chi+N)} \to 0$ for the boundary conditions $cu_{-1}$ and $cu_0$. For this solution, conditions (\ref{eq15}) are given by the values $\tilde{A}_i = cA_i(\chi)$ and the energy dissipation rate (\ref{eq12c}) becomes $\tilde{D} = c^3D(\chi)$. Taking $c = 2^{4(\chi-\tilde{\chi})}$, one writes $\tilde{\nu}_N = 2^{-4(\tilde{\chi}+N)}$ and the obtained relations can be summarized as 
\begin{equation}
\frac{\tilde{D}^{1/3}}{D^{1/3}(\chi)}
= \frac{\tilde{A}_i}{A_i(\chi)}
= \frac{2^{-4\tilde{\chi}}}{2^{-4\chi}},\quad i = 1,2,3.
\label{eq17}
\end{equation}
The two equalities in Eq.~(\ref{eq17}) yield
\begin{equation}
\tilde{A}_i = \tilde{D}^{1/3}\frac{A_i(\chi)}{D^{1/3}(\chi)},\quad
\tilde{\chi}+\frac{\log_2\tilde{D}}{12} = \chi+\frac{\log_2D(\chi)}{12}.
\label{eq17b}
\end{equation}
By expressing $\chi$ from the second equation and substituting into the first one, we obtain 
\begin{equation}
\tilde{A}_i = \tilde{D}^{1/3}\alpha_i\left(
\tilde{\chi}+\frac{\log_2\tilde{D}}{12}\right),
\label{eq17c}
\end{equation}
where the functions $\alpha_i(\tilde{\chi})$ are given implicitly by the equations
\begin{equation}
\alpha_i = \frac{A_i(\chi)}{D^{1/3}(\chi)},\quad
\tilde{\chi} = \chi+\frac{\log_2D(\chi)}{12}.
\label{eq17d}
\end{equation}
Numerical simulations show that the three functions $\alpha_i(\tilde{\chi})$ are identical up to a shift by $i/3$, i.e., (omitting tildes)
\begin{equation}
\alpha_i(\chi) = \alpha\left(\chi-\frac{i}{3}\right),\quad
i = 1,2,3,
\label{eq17e}
\end{equation}
with the periodic function $\alpha(\chi) = \alpha(\chi+1)$ presented in Fig.~\ref{fig4}b. The property (\ref{eq17e}) can be understood as a result of the symmetry (\ref{eqS5}), which yields $A_i \mapsto A_{i+1}$ and $\chi \mapsto \chi-1/3$ for stationary solutions (\ref{eq16}).

Expressions~(\ref{eq17c}) and (\ref{eq17e}) determine the period-3 asymptotic conditions (\ref{eq6}) for the inviscid stationary solution $U_n(\chi)$ with a given dissipation rate $D$ as 
\begin{equation}
\begin{array}{c}
\displaystyle
U_{3n+i} \to 
a_ik_{3n+i}^{-1/3},\\[10pt] 
\displaystyle
a_i(\chi,D) = \sigma_iD^{1/3}\alpha\left(
\chi-\frac{i}{3}+\frac{\log_2D}{12}\right),\quad
i = 1,2,3,\quad
n \to \infty,
\end{array}
\label{eq18}
\end{equation}
where we also took into account the period-3 sign changes allowed by the symmetry (\ref{eqS3}).
Conditions (\ref{eq18}) are universal, i.e., valid for arbitrary boundary conditions, as we confirmed numerically in Fig.~\ref{fig4}b for a number of tests with random values of $u_{-1}$, $u_0$ and $\chi$. In these tests, we took $N = 6$ corresponding to the inertial interval extended to $n_K = 18$, and plotted $\alpha = D^{-1/3}k_{3n+i}^{1/3}|u_{3n+i}|$ versus $\chi-i/3+(\log_2D)/12$ for $i = 1,2,3$ and $n = 4$.   

We conclude that the viscous mechanism has the global effect on stationary solutions, by determining the periodic state in the inertial interval through the condition (\ref{eq18}).
This effect persists in the limit of vanishing viscosity and depends on the way this limit it taken governed by the parameter $\chi$. This implies the nonuniqueness of the vanishing viscosity limit, generating a one-parameter family of inviscid solutions, where 
each solution is defined by a specific viscosity sequence, $\nu_N = 2^{-4(\chi+N)} \to 0$ with fixed $\chi\,(\mathrm{mod}\,1)$. 

\section{Spontaneous stochasticity in the inviscid limit}
\label{sec5}

The boundary condition (\ref{eq18}) was obtained by considering the limit (\ref{eq12}), 
where the limiting expression represents the symmetry transformation (\ref{eqS5}) applied $3N$ times to a 
stationary solution. For time-dependent solutions, this transformation yields $t \mapsto 2^{2N}t$. This means that the time scale increases with $N$ and, hence, the relaxation time of the time-dependent solution to a stationary one vanishes in the limit $N \to \infty$. Therefore, the asymptotic state (\ref{eq18}) must be attained instantaneously in the limit of vanishing viscosity $\nu_N = 2^{-4(\chi+N)} \to 0$. 

\begin{figure}
\centering
\includegraphics[width = 0.45\textwidth]{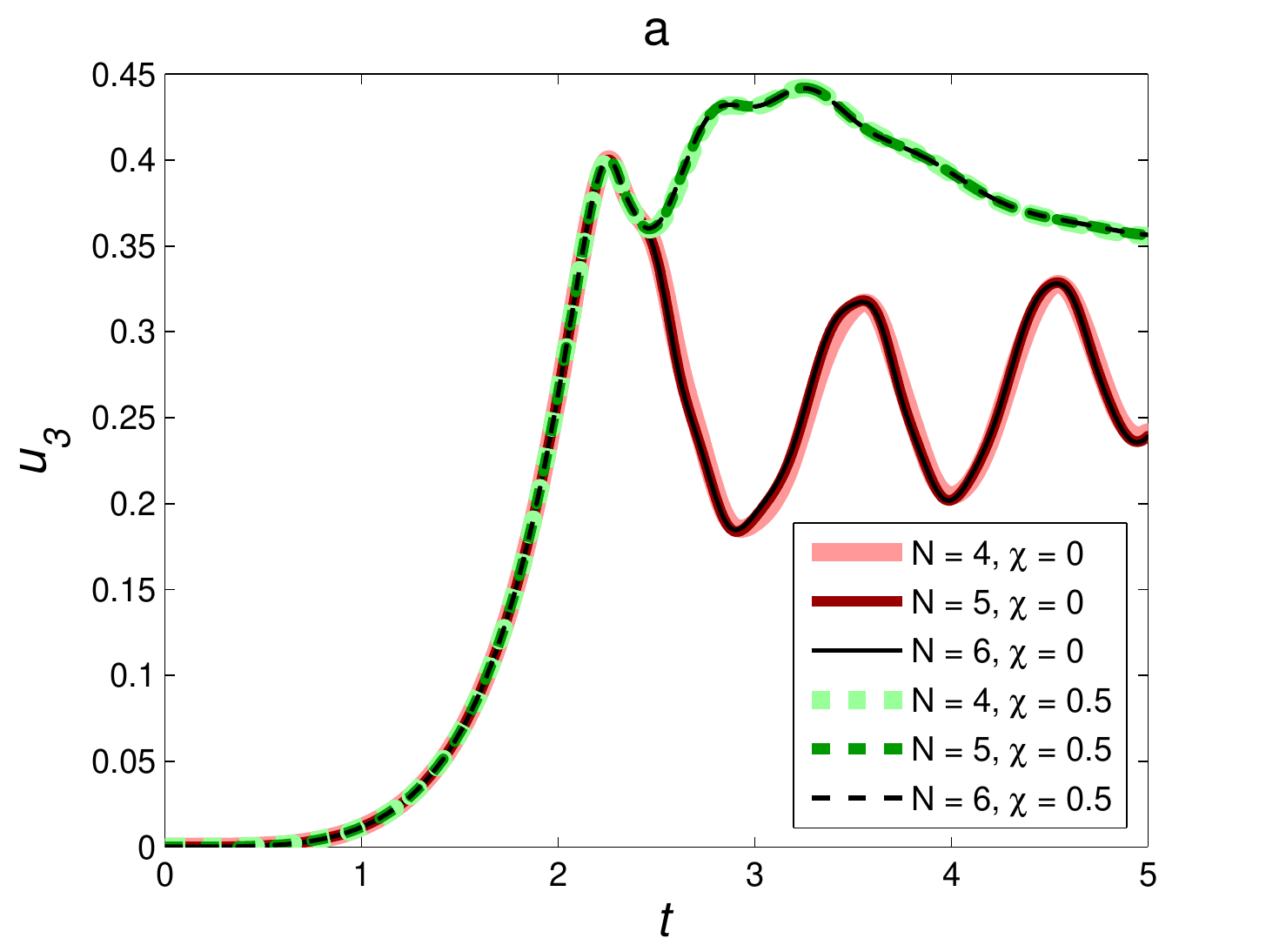}
\includegraphics[width = 0.45\textwidth]{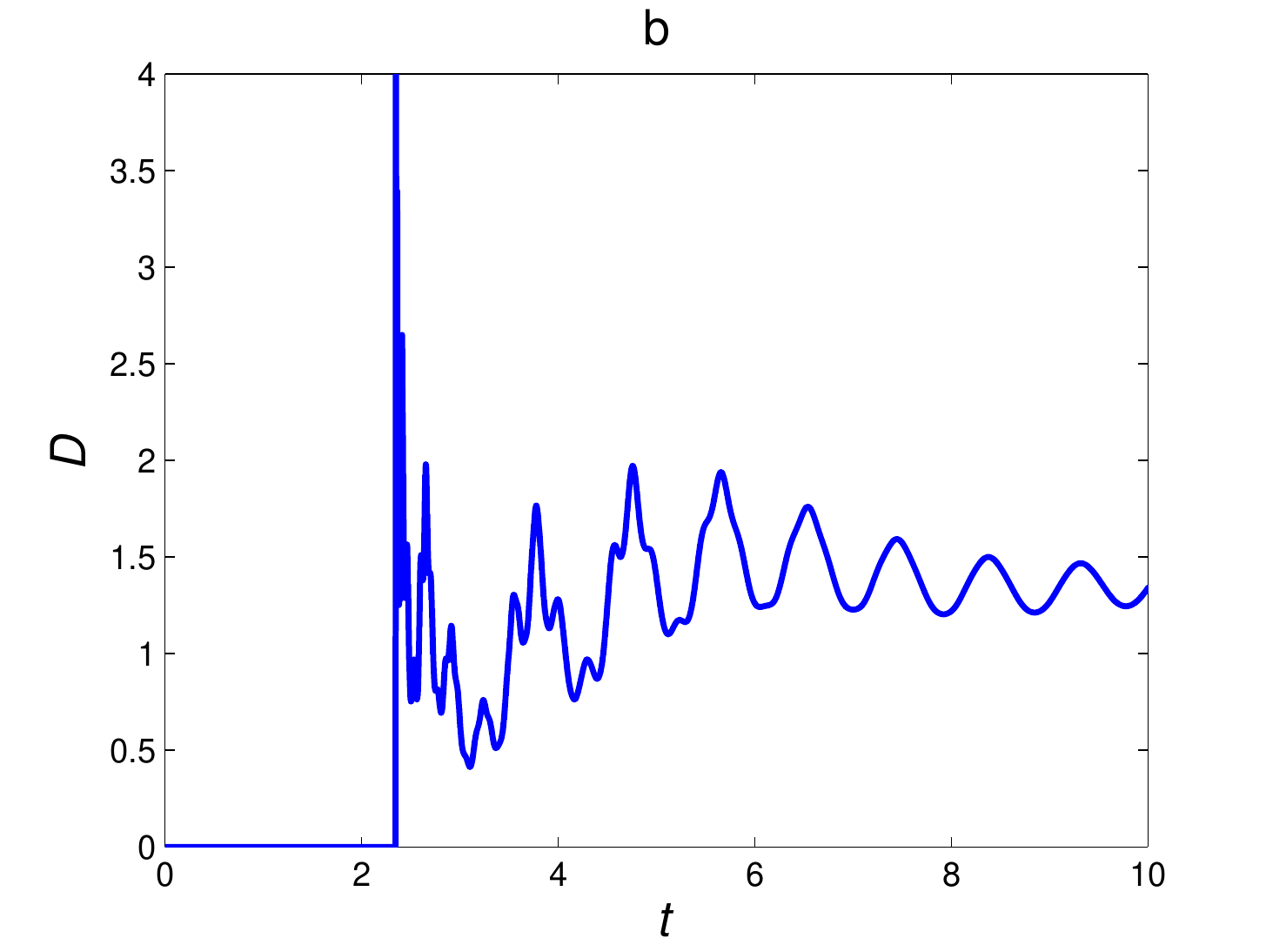}
\includegraphics[width = 0.45\textwidth]{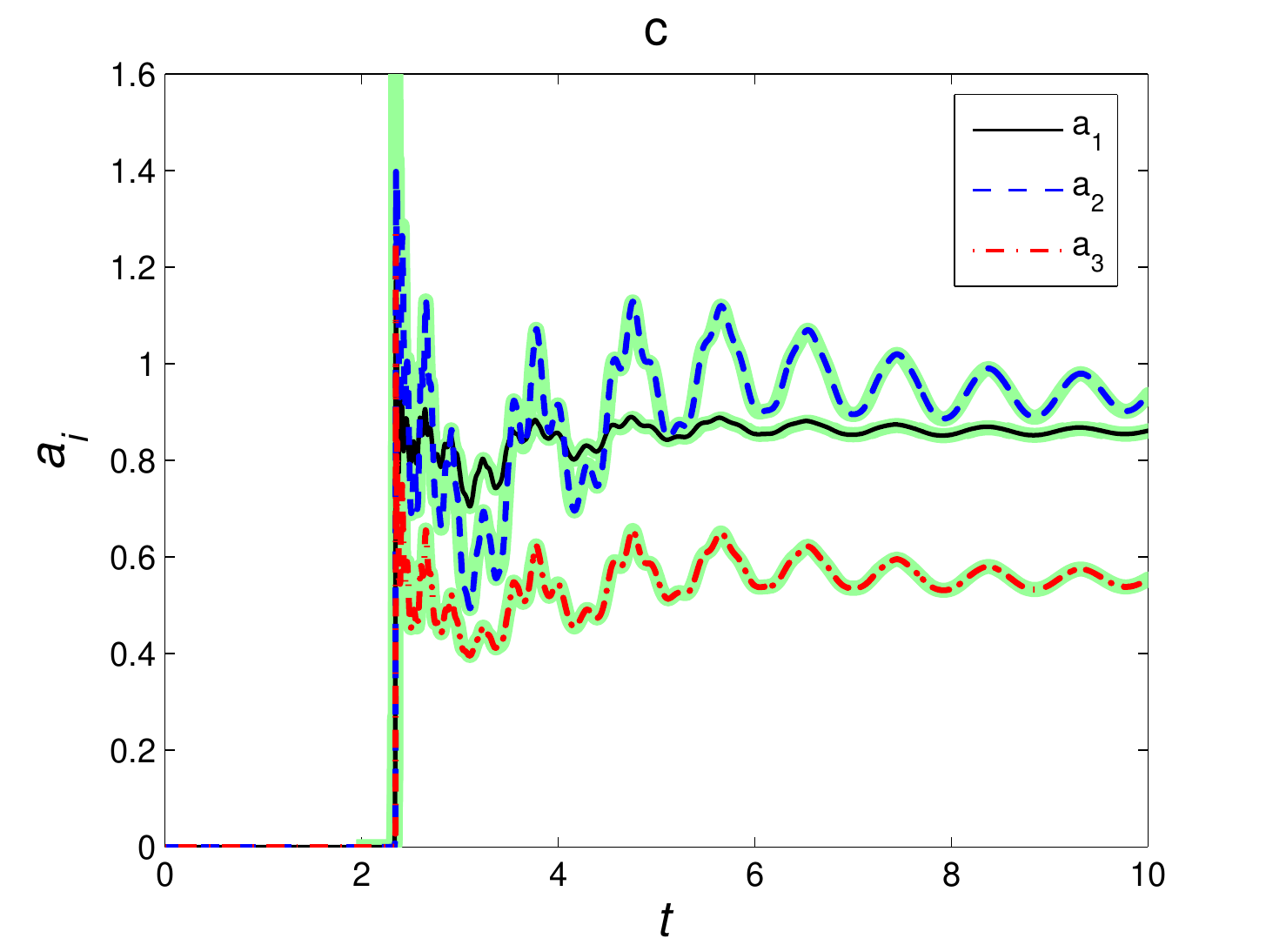}
\caption{a) Solutions for the shell $u_3(t)$ with zero initial conditions and boundary conditions $u_{-1} = u_0 = 0.7$ corresponding to different values of the viscosity $\nu_N = 2^{-4(\chi+N)}$. The limiting inviscid solution depends on the parameter $\chi$  after the blowup time, $t > t_b \approx 2.35$. b) Dissipation rate $D(t)$ as a function of time for the solution with $\chi = 0$ and $N = 6$. c) Values of 
the coefficients $a_i$ determining the solution (\ref{eq20}) in the inertial interval. These coefficient coincide with their prediction based on the value of the dissipation rate (bold light-green lines) given by Eq.~(\ref{eq18}).}
\label{fig5}
\end{figure}

In order to formalize this condition, let us consider a time-dependent solution $u_n^{[\nu]}(t)$ defined for viscosity $\nu > 0$ and fixed initial and boundary conditions. Given the parameter $\chi\,(\mathrm{mod}\,1)$, we consider the vanishing viscosity limit as
\begin{equation}
u_n(t,\chi) = \lim_{N \to \infty}u_n^{[\nu_N]}(t),\quad
\nu_N = 2^{-4(\chi+N)}.
\label{eq19}
\end{equation}
Then, the asymptotic condition (\ref{eq18}) should hold at each time (we will see later that this excludes the time of blowup) as
\begin{equation}
u_{3n+i}(t,\chi) \to a_i(\chi,D(t))k_{3n+i}^{-1/3},\quad
i = 1,2,3,\quad
n \to \infty,
\label{eq20}
\end{equation}
with $a_i(\chi,D)$ given in Eq.~(\ref{eq18}) and the dissipation rate $D(t)$ depending on time.  
Existence of the limit (\ref{eq19}) is confirmed numerically and the condition (\ref{eq20})  
is verified in Fig.~\ref{fig5}. 

Condition (\ref{eq20}) implies that the vanishing viscosity limit, $\nu \to +0$, of time-dependent solutions does not exist in the usual 
sense. However, this  limit can be 
defined for specific sequences of vanishing viscosities, $\nu_N \to 0$. 
This generates an infinite number of the inviscid solutions parameterized by $\chi\ (\mathrm{mod}\,1)$. One can interpret such solutions in different ways.
Since the viscosity $\nu$ has a specific physical value, which can be very small, 
the parameter $\chi = -\frac{1}{4}\log_2\nu\ (\mathrm{mod}\, 1)$ is determined uniquely. Hence, one can use this parameter for considering an appropriate limiting sequence $\nu_N = 2^{-4(\chi+N)} \to 0$. Alternatively, an approximate solution $u_n(t)$ for small $n$ can be found by using the viscosity $\nu_N$ with a relatively small $N$ keeping the same value of $\chi$, i.e., one can obtain a good approximation for shell velocities at large scales without resolving all small scales of the model. This is analogous to the approach in numerical fluid dynamics known as the Large Eddy Simulation, which resolves some but not all of the turbulence scales.

A different interpretation is obtained by  
considering the viscosity to be small but otherwise undetermined (unknown). In this case, the vanishing viscosity limit may be defined by introducing a probability distribution for the viscous parameter $\chi$ as follows. Let us fix the boundary and initial 
conditions and consider $\chi = X$ as a random variable uniformly distributed in the interval $0 \le X \le 1$ (a random value, when chosen, is used for all times $t > 0$). 
Then the inviscid limit is defined as
\begin{equation}
U_n(t) = \lim_{\mu \to +0}u_n^{[\nu]}(t),\quad
\nu = \mu 2^{-4X}.
\label{eq21}
\end{equation}
At each time, the solution $u_n^{[\nu]}(t)$, $n = 1,2,\ldots$, represents 
a random variable (probability measure) in the $\ell^2$ space with the norm given by the square root of energy, and the limit can be understood in a weak sense. 
Note that no choice of a 
special viscosity subsequence is necessary in the limit (\ref{eq21}), where all limiting solutions (\ref{eq19}) are involved through the random variable $X$. As a result, the inviscid solution $U_n(t)$ is given 
by a singular probability measure supported on the one-parameter set of solutions (\ref{eq19}). 

\begin{figure}
\centering
\includegraphics[width = 1\textwidth]{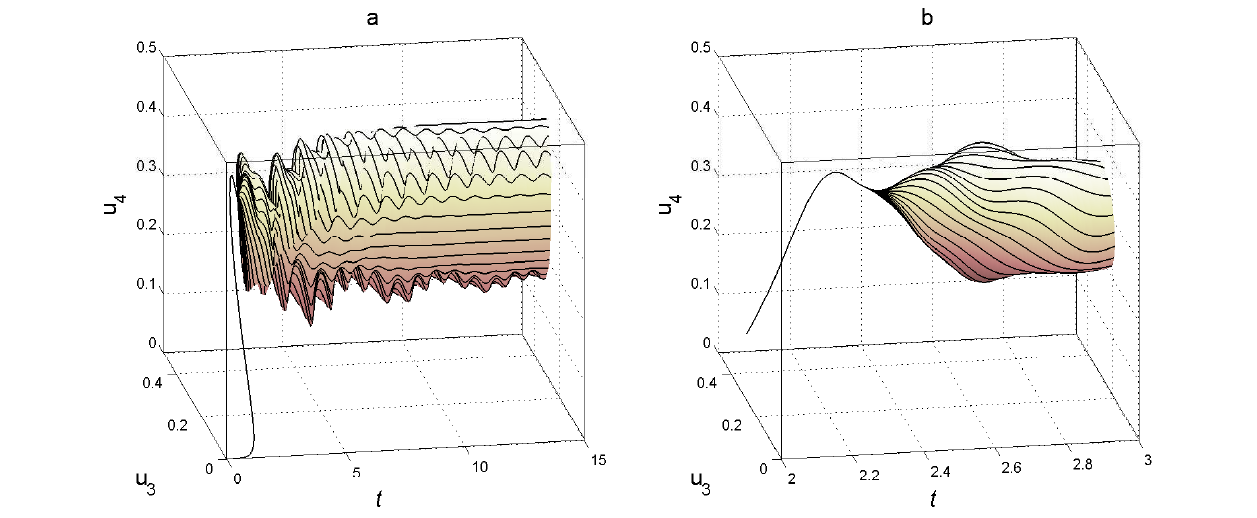}
\caption{Evolution of the shells $(u_3,u_4)$ for the solution in vanishing viscosity limit for zero initial 
conditions and boundary values $u_{-1} = u_0 = 0.7$. The curves show solutions for very small 
viscosities $\nu = 2^{-4(\chi+6)}$ with $\chi = 0,0.04,\ldots,1$. The inviscid 
solutions coincide for all $0 \le \chi < 1$ before the blowup, $t \le t_b$, and define a surface for larger times, $t > t_b$.  
The right figure shows the amplified region near the blowup time $t_b \approx 2.35$, when the spontaneous stochasticity occurs.}
\label{fig6}
\end{figure}

Figure~\ref{fig6} shows the solution (\ref{eq21}) computed numerically.
One can see that the limiting solution 
is deterministic until a certain time $t \le t_b$ 
(with the blowup time $t_b$ as described in the next section), and becomes stochastic  for $t > t_b$. 
This reveals the striking property of the spontaneous stochasticity of the inviscid solution $U_n(t)$ 
obtained in the limit of vanishing viscosity. 

\begin{figure}
\centering
\includegraphics[width = 0.55\textwidth]{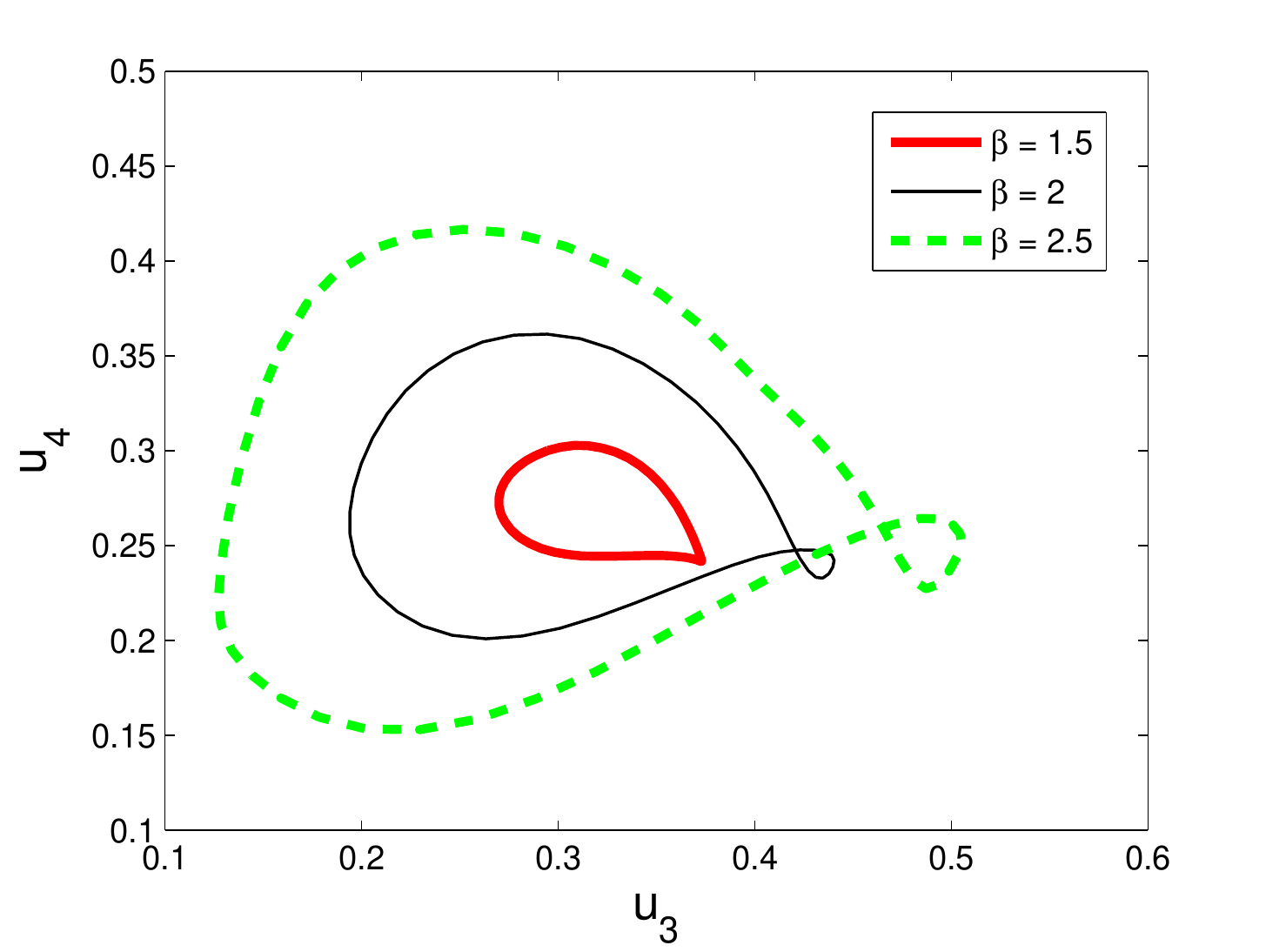}
\caption{Shell speeds $(u_3,u_4)$ at time $t = 3$ of the vanishing viscosity solutions. Different viscosity mechanisms are considered given by the term $-\nu k_n^{\beta}u_n$ in Eq.~(\ref{eq1}) for $\beta = 1.5$, $2$ and $2.5$. The results are obtained numerically for zero initial conditions and boundary values $u_{-1} = u_0 = 0.7$. The presented curves are determined by the viscosity spanning the interval $0 < \nu \le \nu_0$ with very small $\nu_0$, such that the viscous range starts around $n_K \approx 24$. Inviscid solutions can be represented as singular probability measures (supported on a curve for the shells $u_3$ and $u_4$), which depend on the infinitesimal viscosity mechanism.}
\label{fig6b}
\end{figure}

Figure~\ref{fig6b} (thin solid line) shows the one-dimensional support of the singular probability measure (\ref{eq21}) for the shell speeds $u_3$ and $u_4$ computed numerically at $t = 3$. This support represents a closed curve. It is remarkable that the probability measure depends on the viscosity mechanism. Indeed, let us consider Eq.~(\ref{eq1}) with the hyperviscous term $-\nu k_n^{\beta}u_n$, where the usual viscosity corresponds to $\beta = 2$. The general case with $\beta \ne 2$ can be studied in a similar way, where one should use a different scaling of viscosity depending on $\beta$. The inviscid solution in the vanishing viscosity limit can be defined as a probability measure, similarly to Eq.~(\ref{eq21}). This measure was computed numerically for $\beta = 1.5$ and $2.5$, see Fig.~\ref{fig6b} (thick solid and dashed lines). The simulations confirm that the measure is singular with a one-dimensional support, which is different for different $\beta$, i.e., the inviscid limit depends strongly on the viscosity mechanism. 

\section{Blowup} 
\label{sec6}

Let us consider the Cauchy problem for the system (\ref{eq1}) with arbitrary boundary conditions (\ref{eq1b}) and initial conditions at $t = 0$ with finite enstrophy $\Omega(0) = \sum k_n^2u_n^2 < \infty$. In the viscous case, $\nu > 0$, there exists a unique solution~\cite{barbato2006some,constantin2006analytic,constantin2007regularity}. For the inviscid system, $\nu = 0$, the solution exists in a weak sense and it is unique as soon as the enstrophy is finite. The enstrophy may explode (blowup) in finite time $t_b > 0$ such that $\Omega \to \infty$ as $t \to t_b^-$. 
Before the blowup, $0 \le t < t_b$, the viscous solutions converge to the inviscid one in the vanishing viscosity limit $\nu \to +0$~\cite{barbato2006some,constantin2006analytic,constantin2007regularity}. 
As we have shown in Section~\ref{sec5}, the limit $\nu \to +0$ does not exist and the inviscid solution defined for vanishing viscosity subsequences is not unique for larger times.

The blowup structure at times $t < t_b$ was analyzed in~\cite{dombre1998intermittency} 
(see also~\cite{mailybaev2012,mailybaev2012c}) showing that, at large shell numbers $n$, the solution has 
the self-similar asymptotic form
\begin{equation}
u_n(t) \to \sigma_n c k_n^{-y}U(c\xi)
\quad \textrm{with}
\quad
\xi = k_n^{1-y}(t-t_b) \le 0.
\label{eq22}
\end{equation}
Here the scaling factor $c > 0$ and the signs $\sigma_n = \pm 1$ describe 
the symmetries (\ref{eqS2}) and (\ref{eqS3}). The exponent $y \approx 0.281$ is universal 
(independent of boundary and initial conditions). The function $U(\xi)$ is also universal and 
defined for $\xi \le 0$ corresponding to $t \le t_b$, with $U(0) = 1$. 
The convergence in Eq.~(\ref{eq22}) is pointwise for large $n$ and arbitrary fixed $\xi$, i.e., 
for 
\begin{equation}
t = t_b+\xi k_n^{y-1} \to t_b^{-},\quad n \to \infty.
\label{eq22b}
\end{equation}

\begin{figure}
\centering
\includegraphics[width = 1\textwidth]{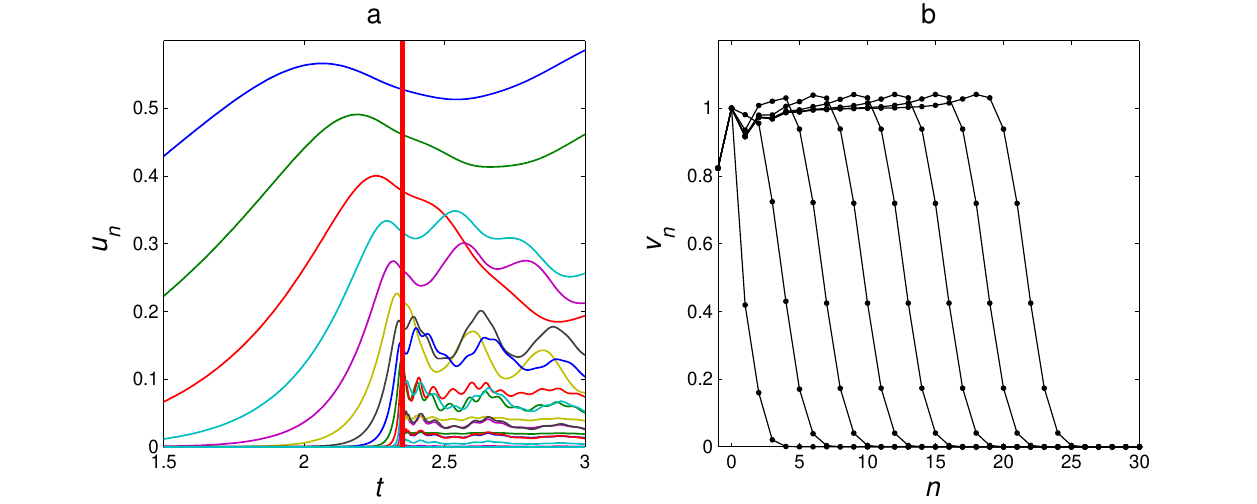}
\caption{a) Magnified version of Fig.~\ref{fig1} near the blowup time $t_b \approx 2.35$ represented by the vertical red line. b) Before the blowup, $t < t_b$, the self-similar inviscid dynamics (\ref{eq22}) corresponds to the asymptotic traveling wave (\ref{eq23}) in renormalized variables $v_n(\tau)$ given by Eq.~(\ref{eq23c}). Shown are the variables $v_n$, $n = -1,0,\ldots,30$, at logarithmic times $\tau = 0,3\tau_0,\ldots,21\tau_0$ for the inviscid solution with zero initial 
conditions and boundary values $u_{-1} = u_0 = 0.7$. The symmetry parameters in Eq.~(\ref{eq23c}) are $c = 0.7$ and $\sigma_n = +1$.}
\label{fig7}
\end{figure}

In our example in Fig.~\ref{fig7}a, the blowup time of the inviscid system is $t_b = 2.35$ and one can recognize the self-similar behavior (\ref{eq22}) for $t < t_b$. At the blowup time, one has $\xi = 0$ in Eq.~(\ref{eq22}), 
which implies the asymptotic power law
\begin{equation}
u_n(t_b) \to \sigma_nck_n^{-y}\quad 
\textrm{as}\quad 
n \to \infty.
\label{eq24}
\end{equation}
Note that, since $y < 1/3$, the energy flux (\ref{eq1d}) tends to infinity for large shell numbers 
at the blowup time $t_b$, 
while the energy is finite, $E(t_b) < \infty$, and the enstrophy is infinite, $\Omega(t_b) = \infty$, 
see also Fig.~\ref{fig5}b.

Let us introduce the new shell variables as
\begin{equation}
u_n = \sigma_nck_n^{-y}v_n,
\quad 
t = t_b-2^{-\tau}/c.
\label{eq23c}
\end{equation}
Then, following \cite{dombre1998intermittency,mailybaev2012}, expression (\ref{eq22}) can be written 
in the form 
\begin{equation}
v_n(\tau) \to V\left(n-\frac{\tau}{\tau_0}\right),
\label{eq23}
\end{equation}
where the function $V(\eta)$, $\eta \in \mathbb{R}$, and constant $\tau_0$ are defined as
\begin{equation}
V(\eta) = U\left(-2^{\tau_0\eta}\right),\quad
\tau_0 = 1-y  \approx 0.719.
\label{eq23b}
\end{equation}
According to Eqs.~(\ref{eq22b}) and (\ref{eq23b}), the convergence in Eq.~(\ref{eq23}) is understood in the limit
\begin{equation}
\tau = \tau_0n+const \to \infty,
\quad n \to \infty,
\label{eq23e}
\end{equation}
i.e., pointwise for a constant $\eta = n-\tau/\tau_0$.

The function $V(\eta)$ has the limits
\begin{equation}
\lim_{\eta \to -\infty}V(\eta) = 1,\quad
\lim_{\eta \to \infty}V(\eta) = 0,
\label{eq23d}
\end{equation}
where the first condition follows from the property $U(0) = 1$. In the second condition, large $\eta$ corresponds to the region of large shell numbers, where $u_n$ decays faster than $k_n^{-1}$ due to the finite enstrophy condition and, hence, $v_n \to 0$ in Eq.~(\ref{eq23c}).

Note that the limit $\tau \to \infty$ for the logarithmic time $\tau = -\log_2[c(t_b-t)]$ corresponds to $t \to t_b^-$. 
Hence, expression (\ref{eq23}) describes the blowup as a traveling wave with the universal stationary profile $V(\eta)$ 
moving from smaller to larger shell numbers with constant speed $\tau_0^{-1}$ in logarithmic time $\tau$, see Fig.~\ref{fig7}b. 
This implies periodicity of the rescaled shell speeds $v_n(\tau)$ in Eq.~(\ref{eq23}), which attain the same values with the shift by one shell number after each period $\tau_0$.   

\section{Onset of spontaneous stochasticity after the blowup} 
\label{sec7}

After the blowup, $t > t_b$, the inviscid solution is not unique, as we observed in numerical simulations, Figs.~\ref{fig5}a and~\ref{fig6}b.
We showed that this nonuniqueness is characterized by the parameter $\chi\, (\mathrm{mod}\,1)$ in Eq.~(\ref{eq19}), determining a specific form of the inviscid limit, 
$\nu_N = 2^{-4(\chi+N)} \to 0$. At large times, the solutions $u_n(t,\chi)$ converge to the stationary ones, 
as we explained in the previous sections. In this section, we focus on the behavior just after the blowup, i.e., 
in the limit $t \to t_b^+$, describing the onset of nonuniqueness and, thus, spontaneous stochasticity. 

Similarly to Eqs.~(\ref{eq22})--(\ref{eq23c}), we introduce the functions $w_n(\tilde{\tau},\chi)$ by changing the variables as
\begin{equation}
u_n = \sigma_n c k_n^{-y}w_n,
\quad
t = t_b+2^{-\tilde{\tau}}/c,
\label{eq25c}
\end{equation}
where $c > 0$ and the signs $\sigma_n$ describe the symmetry transformations (\ref{eqS2}) and (\ref{eqS3}). The second expression introduces a new logarithmic time variable $\tilde{\tau} = -\log_2[c(t-t_b)]$ after the blowup, $t > t_b$, with the right blowup limit, $t \to t_b^+$, 
corresponding to large $\tilde{\tau} \to \infty$. 
An arbitrary solution $w_n(\tilde{\tau},\chi)$ can be represented as
\begin{equation}
\begin{array}{c}
\displaystyle
w_n(\tilde{\tau},\chi) = 
W_n\left(n-\frac{\tilde{\tau}}{\tau_0},
\chi-\chi_c-\frac{\tilde{\tau}}{\tau_1}\right),
\\[15pt]
\displaystyle
\chi_c = -\frac{1}{4}\log_2c,
\quad
\tau_1 = \frac{4-4y}{1+y} \approx 2.245,
\end{array}
\label{eq25d}
\end{equation}
where $\tau_0 = 1-y$ as in Eq.~(\ref{eq23b}); the constant $\chi_c$ is induced by the change of viscosity $\nu_N = 2^{-4(\chi+N)}$ due to symmetry transformation (\ref{eqS2}) with $c = 2^{-4\chi_c}$. This representation resembles Eq.~(\ref{eq23}), but takes into account the dependence on $\chi$. The reason for the choice of the second arguments in $W_n$ will be clear below. 

It is straightforward to check that the transformation
\begin{equation}
u_n \mapsto \sigma_{n-1}\sigma_n 2^{-y}u_{n-1},\quad t-t_b\mapsto 2^{y-1}(t-t_b), \quad 
\nu \mapsto 2^{-(1+y)}\nu
\label{eqS4}
\end{equation}
is the symmetry, i.e., it relates different solutions of Eq.~(\ref{eq1}). 
It is easy to see that the asymptotic form (\ref{eq24}) 
of the inviscid solution at the blowup point ($t = t_b$ and $\nu = 0$) does not change under the transformation (\ref{eqS4}). For inviscid solutions $u_n(t,\chi)$, the symmetry (\ref{eqS4}) transforms the new variables (\ref{eq25c}) as 
\begin{equation}
w_n \mapsto w_{n-1},\quad
\tilde{\tau} \mapsto \tilde{\tau}+\tau_0
\label{eqS4c}
\end{equation}
with $\tau_0 = 1-y$. Also, the last relation in Eq.~(\ref{eqS4}) yields the mapping 
\begin{equation}
\chi \mapsto \chi+\chi_0,\quad \chi_0 = \frac{1+y}{4} = \frac{\tau_0}{\tau_1}
\label{eqS4b}
\end{equation}
for the parameter of the vanishing viscosity limit $\nu_N = 2^{-4(\chi+N)} \to 0$ with $\tau_1$ given by Eq.~(\ref{eq25d}). We see from Eqs.~(\ref{eqS4c}) and (\ref{eqS4b}) that the arguments $\eta_1 = n-\tilde{\tau}/\tau_0$ and $\eta_2 = \chi-\chi_c-\tilde{\tau}/\tau_1$ in Eq.~(\ref{eq25d}) are chosen such that the functions $W_n$ transform simply as
\begin{equation}
W_n(\eta_1,\eta_1) \mapsto W_{n-1}(\eta_1,\eta_2),
\label{eqS4d}
\end{equation}
i.e., only with the change of shell number.

Recall that the asymptotic state (\ref{eq24}) of the system at the blowup point $t_b$ is universal, i.e., it is independent of initial and boundary conditions up to the choice of the scaling parameter $c$ and signs $\sigma_n$. Hence, we can expect that similar universality holds after the blowup as well. Thus, we conjecture (and confirm later numerically) that the functions in Eq.~(\ref{eq25d}) have a universal asymptotic form for large $n$. This asymptotic form should not be affected by the symmetry transformation (\ref{eqS4}), which leaves the asymptotic state (\ref{eq24}) unchanged. Since this transformation changes the shell number by one in the functions (\ref{eqS4d}), their universal asymptotic form, $W_n(\eta_1,\eta_2) \to W(\eta_1,\eta_2)$, must be the independent of the shell number $n$. Using Eq.~(\ref{eq25d}), this yields the asymptotic expression of the form
\begin{equation}
w_n(\tilde{\tau},\chi) \to
W\left(n-\frac{\tilde{\tau}}{\tau_0},
\chi-\chi_c-\frac{\tilde{\tau}}{\tau_1}\right).
\label{eqS4e}
\end{equation}
Similarly to Eqs.~(\ref{eq23}) and (\ref{eq23e}), we understand the limit (\ref{eqS4e}) pointwise for large $n \to \infty$ with fixed $\eta_1 = n-\tilde{\tau}/\tau_0$ and $\eta_2 = \chi-\chi_c-\tilde{\tau}/\tau_1$. 

Recall that the values of $\chi$, which differ by an integer number, correspond to the same inviscid solution, see Section~\ref{sec5}. Hence, the universal function $W(\eta_1,\eta_2)$ is periodic with respect to the second variable as
\begin{equation}
W(\eta_1,\eta_2) = W(\eta_1,\eta_2+1).
\label{eq35}
\end{equation}
The power law (\ref{eq24}), where $t_b$ corresponds to $\tilde{\tau} \to \infty$, and the relations (\ref{eq25c}), (\ref{eqS4e}), (\ref{eq35}) yield
the left-side limiting value of the function $W(\eta_1,\eta_2)$ with respect to the first argument as  
\begin{equation}
\lim_{\eta_1 \to -\infty} W(\eta_1,\eta_2) = 1.
\label{eq26}
\end{equation}
The limit on the other side follows from the period-3 condition (\ref{eq20}), which implies that $w_n \sim k_n^{y-1/3} \to 0$ for large $n$ with $y-1/3 < 0$. This yields
\begin{equation}
\lim_{\eta_1 \to \infty} W(\eta_1,\eta_2) = 0.
\label{eq27}
\end{equation}
Note that the convergence in Eq.~(\ref{eq27}) is rather slow due to the small absolute value of the exponent $y-1/3 \approx -0.0524$.
   
\begin{figure}
\centering
\includegraphics[width = 1\textwidth]{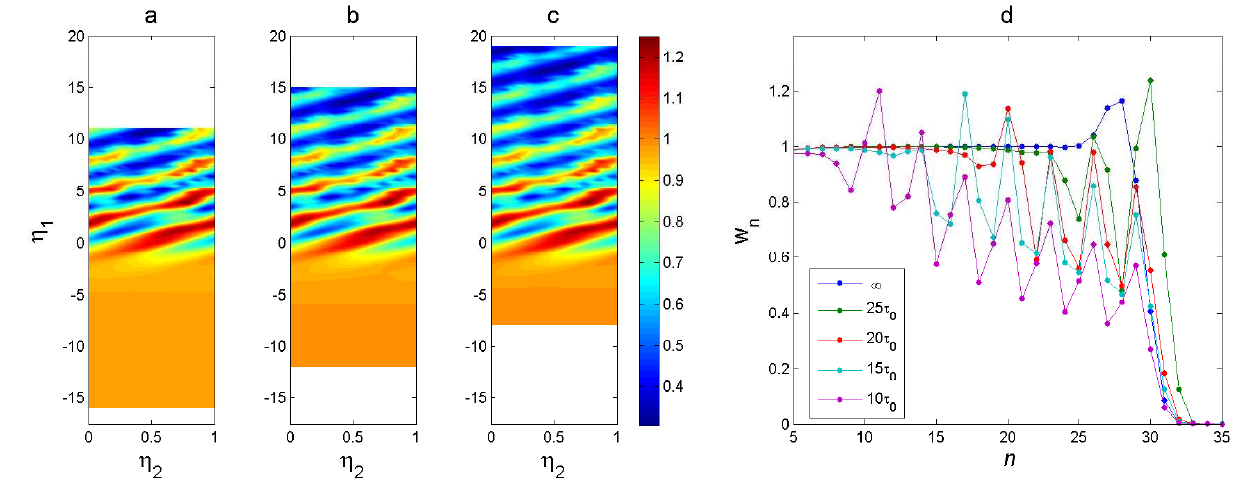}
\caption{a,b) Functions $W_{14}$ and $W_{18}$ describing behavior after the blowup in new shell coordinates and logarithmic time (\ref{eq25d}). The functions are computed numerically using Eq.~(\ref{eq1}) with $35$ shells for viscosities $\nu = 2^{-4(\chi+10)}$, $\chi = 0,0.01,\ldots,0.99$. The boundary and initial conditions are as in Fig.~\ref{fig1}. c) Similar computations for the function $W_{22}$ but for the boundary and initial conditions $u_n = k_n^{-y}$. The functions in figures a-c are almost identical, confirming existence of the universal asymptotic form (\ref{eqS4e}). Note that these figures correspond to the same interval $3\tau_0 \le \tilde{\tau} \le 30\tau_0$, which yields different intervals of $\eta_1 = n-\tilde{\tau}/\tau_0$ for different $n$ in Eq.~(\ref{eq25d}).
d) Renormalized shell variables $w_n(\tilde{\tau},\chi)$ computed under conditions of figures a,b for $\chi = 0$ and different $\tilde{\tau}$, demonstrating quasi-periodic dynamics.}
\label{fig8}
\end{figure}

The conjectured universal asymptotic form (\ref{eqS4e}) fully agrees with the numerical simulations. In Fig.~\ref{fig8}a,b we show the results of high-precision simulations carried out for Eq.~(\ref{eq1}) with the boundary conditions $u_{-1} = u_0 = 0.7$, zero initial conditions and very small viscosities $\nu = 2^{-4(\chi+10)} \sim 10^{-13}$ with $\chi = 0,0.01,\ldots,0.99$ (the viscous range starts at $n_K \approx 30$). Shown are the functions $W_n(\eta_1,\eta_2)$ for $n = 14$ and $18$, which are determined by the corresponding shell speeds using Eqs.~(\ref{eq25c}) and (\ref{eq25d}). These functions appear to be almost identical, confirming the asymptotic relation (\ref{eqS4e}). Figure~\ref{fig8}c shows the function $W_{22}(\eta_1,\eta_2)$ of the analogous simulation, but for the boundary and initial conditions $u_n = k_n^{-y}$, $n = -1,0,1,\ldots$. These conditions correspond to the power law (\ref{eq24}) at the blowup point satisfied exactly. Since the function $W_{22}$ is the same as the functions $W_{14}$ and $W_{18}$ in Fig.~\ref{fig8}a,b, we confirmed the universality of the asymptotic form (\ref{eqS4e}), i.e., its independence of the boundary and initial conditions. 
Note that relations (\ref{eq35})--(\ref{eq27}) are clearly satisfied in Fig.~\ref{fig8}a-c with very slow convergence in the last condition, as it was expected. 

The asymptotic form (\ref{eqS4e}) contains different periods $\tau_0 \approx 0.719$ and $\tau_1 \approx 2.245$ in two different arguments of the function $W$. This means that, in renormalized variables (\ref{eq25c}), the dynamics after the blowup is quasi-periodic (together with the shift by one shell number after each $\tau_0$), unlike the periodic dynamics before the blowup in Eq.~(\ref{eq23}), see Figs.~\ref{fig7} and \ref{fig8}. Figure~\ref{fig8}d shows this quasi-periodic dynamics with the logarithmic time intervals of $5\tau_0$. Here the viscous range starts at $n_K = 30$, so that the formula (\ref{eqS4e}) for the inviscid dynamics is appropriate for the shells $n \lesssim 25$.  As $\tilde{\tau}$ decreases from infinity (the original time $t$ increases in the region $t > t_b$), the lower shell numbers get involved in the dynamics, while the higher shell numbers develop the period-3 tail as described in Section~\ref{sec5}.

We conclude that the self-similar wave (\ref{eq23}) propagating to large shell numbers (from large to small scales) before the blowup is followed 
by the wave (\ref{eqS4e}) moving in the opposite direction (from small to large scales) after the blowup. The latter wave is given by a periodic function (\ref{eq35}), which is universal (independent of initial and boundary conditions) and contains the information about all inviscid solutions 
obtained in the limit of vanishing viscosity.
This completes the description for the onset of spontaneous stochasticity, which occurs after the blowup, see Fig.~\ref{fig6}b. 

\section{Conclusion}

In this work, we provided the detailed numerical and theoretical description of the spontaneous stochasticity phenomenon in the real version of the GOY shell model of turbulence. This deterministic model describes the energy transport from large to small scales, which is qualitatively similar to the one observed in the developed turbulence for the 3D Navier-Stokes equations. Essential reduction of the number of degrees of freedom for the shell model allows high-accuracy multiscale numerical analysis. We demonstrated that the limit of vanishing viscosity, chosen in a specific form with a single real parameter, is not unique and leads to an infinite number of different inviscid solutions. Existence of stable stationary solutions in the model allows understanding the inviscid limit both for stationary and time-dependent solutions, where infinitesimal viscosity strongly affects the system at all scales. We showed that the inviscid solution, which is unique until the finite-time blowup, becomes nonunique after the blowup. Emergence of the infinite number of solutions is described using the renormalization technique. This approach shows that the inviscid solutions follow a universal quasi-periodic (in renormalized coordinates) asymptotic law triggered by the infinitesimal viscosity, with a perturbation propagating from infinitely small to large scales in finite time.

Considering the vanishing viscosity limit in probabilistic sense, with the viscosity as a small random variable, one can define the unique inviscid solution as a probability measure. We showed that this probability measure depends on the viscous mechanism: for example, it is different for a system with hyperviscosity. We stress the fact that the stochastic solution appears spontaneously in finite time for an otherwise deterministic system (with no stochastic forcing) and for deterministic initial and boundary conditions.  These results indicate the way for possible formalization of the spontaneous stochasticity phenomenon, which may occur in systems with finite-time blowup. Note that the probability measure in our solution is singular, as it involves a one-dimensional set of solutions. In a forthcoming work, we give the numerical evidence that a similar approach leads to a continuous probability measure when applied to chaotic systems, i.e., the inviscid limit yields a spontaneously stochastic solution, which is continuously distributed in the infinite-dimensional configuration space.

\section*{Acknowledgments} 
The work was supported by the CNPq (grant 305519/2012-3) and by the FAPERJ
(Pensa Rio 2014).

\bibliographystyle{plain}
\bibliography{refs}

\begin{thebibliography}{10}

\bibitem{agafontsev2015}
D.S. Agafontsev, E.A. Kuznetsov, and A.A. Mailybaev.
\newblock {Development of high vorticity structures in incompressible 3D Euler
  equations}.
\newblock {\em arXiv:1502.01562}, 2015.

\bibitem{barbato2006some}
D.~Barbato, M.~Barsanti, H.~Bessaih, and F.~Flandoli.
\newblock {Some rigorous results on a stochastic GOY model}.
\newblock {\em J. Stat. Phys.}, 125(3):677--716, 2006.

\bibitem{biferale2003shell}
L.~Biferale.
\newblock Shell models of energy cascade in turbulence.
\newblock {\em Annu. Rev. Fluid Mech.}, 35(1):441--468, 2003.

\bibitem{biferale1995transition}
L.~Biferale, A.~Lambert, R.~Lima, and G.~Paladin.
\newblock Transition to chaos in a shell model of turbulence.
\newblock {\em Physica D}, 80(1):105--119, 1995.

\bibitem{bustamante20083d}
M.~D. Bustamante and R.~M. Kerr.
\newblock {3D Euler about a 2D symmetry plane}.
\newblock {\em Physica D}, 237(14):1912--1920, 2008.

\bibitem{cheskidov2007inviscid}
A.~Cheskidov, S.~Friedlander, and N.~Pavlovi{\'c}.
\newblock {Inviscid dyadic model of turbulence: the fixed point and Onsager’s
  conjecture}.
\newblock {\em Journal of Mathematical Physics}, 48(6):065503, 2007.

\bibitem{constantin2006analytic}
P.~Constantin, B.~Levant, and E.S. Titi.
\newblock Analytic study of shell models of turbulence.
\newblock {\em Physica D}, 219(2):120--141, 2006.

\bibitem{constantin2007regularity}
P.~Constantin, B.~Levant, and E.S. Titi.
\newblock Regularity of inviscid shell models of turbulence.
\newblock {\em Phys. Rev. E}, 75(1):016304, 2007.

\bibitem{desnyansky1974evolution}
V.N. Desnyansky and E.A. Novikov.
\newblock The evolution of turbulence spectra to the similarity regime.
\newblock {\em Izv. A.N. SSSR Fiz. Atmos. Okeana}, 10(2):127--136, 1974.

\bibitem{dombre1998intermittency}
T.~Dombre and J.L. Gilson.
\newblock {Intermittency, chaos and singular fluctuations in the mixed
  Obukhov--Novikov shell model of turbulence}.
\newblock {\em Physica D}, 111(1--4):265--287, 1998.

\bibitem{eyink2015spontaneous}
G.L. Eyink and T.D. Drivas.
\newblock {Spontaneous stochasticity and anomalous dissipation for Burgers
  equation}.
\newblock {\em J. Stat. Phys.}, 158(2):386--432, 2015.

\bibitem{eyink2006onsager}
G.L. Eyink and K.R. Sreenivasan.
\newblock {Onsager and the theory of hydrodynamic turbulence}.
\newblock {\em Rev. Modern Phys.}, 78(1):87--135, 2006.

\bibitem{falkovich2001particles}
G.~Falkovich, K.~Gawedzki, and M.~Vergassola.
\newblock Particles and fields in fluid turbulence.
\newblock {\em Rev. Mod. Phys.}, 73(4):913, 2001.

\bibitem{frisch1995turbulence}
U.~Frisch.
\newblock {\em {Turbulence: The Legacy of A.N. Kolmogorov}}.
\newblock Cambridge University Press, 1995.

\bibitem{frisch1999turbulence}
U.~Frisch.
\newblock {\em {Turbulence: the legacy of A.N.~Kolmogorov}}.
\newblock Cambridge University Press, 1999.

\bibitem{gibbon2008three}
J.~D. Gibbon.
\newblock {The three-dimensional Euler equations: Where do we stand?}
\newblock {\em Physica D}, 237(14-17):1894--1904, 2008.

\bibitem{gilson1997towards}
J.~L. Gilson and T.~Dombre.
\newblock Towards a two-fluid picture of intermittency in shell models of
  turbulence.
\newblock {\em Phys. Rev. Lett.}, 79(25):5002--5005, 1997.

\bibitem{gledzer1973system}
E.B. Gledzer.
\newblock System of hydrodynamic type admitting two quadratic integrals of
  motion.
\newblock {\em Sov. Phys. Doklady}, 18:216, 1973.

\bibitem{grafke2008numerical}
T.~Grafke, H.~Homann, J.~Dreher, and R.~Grauer.
\newblock {Numerical simulations of possible finite time singularities in the
  incompressible Euler equations: comparison of numerical methods}.
\newblock {\em Physica D}, 237(14):1932--1936, 2008.

\bibitem{hou2008blowup}
T.~Y. Hou and R.~Li.
\newblock {Blowup or no blowup? The interplay between theory and numerics}.
\newblock {\em Physica D}, 237(14):1937--1944, 2008.

\bibitem{kolmogorov1941local}
A.N. Kolmogorov.
\newblock {The local structure of turbulence in incompressible viscous fluid
  for very large Reynolds numbers}.
\newblock {\em Dokl. Akad. Nauk SSSR}, 30:9--13, 1941.

\bibitem{landau2013fluid}
L.D. Landau and E.M. Lifshitz.
\newblock {\em {Fluid Mechanics}}.
\newblock Elsevier, 2013.

\bibitem{luo2014}
G.~Luo and T.Y. Hou.
\newblock {Toward the finite-time blowup of the 3D axisymmetric Euler
  equations: a numerical investigation}.
\newblock {\em Multiscale Model. Simul.}, 12:1722--1776, 2014.

\bibitem{l1998improved}
V.S. L'vov, E.~Podivilov, A.~Pomyalov, I.~Procaccia, and D.~Vandembroucq.
\newblock Improved shell model of turbulence.
\newblock {\em Phys. Rev. E}, 58(2):1811, 1998.

\bibitem{mailybaev2012computation}
A.A. Mailybaev.
\newblock Computation of anomalous scaling exponents of turbulence from
  self-similar instanton dynamics.
\newblock {\em Phys. Rev. E}, 86(2):025301, 2012.

\bibitem{mailybaev2012}
A.A. Mailybaev.
\newblock Renormalization and universality of blowup in hydrodynamic flows.
\newblock {\em Phys. Rev. E}, 85(6):066317, 2012.

\bibitem{mailybaev2012c}
A.A. Mailybaev.
\newblock {Bifurcations of blowup in inviscid shell models of convective
  turbulence}.
\newblock {\em Nonlinearity}, 26:1105--1124, 2013.

\bibitem{mailybaev2013blowup}
A.A. Mailybaev.
\newblock Blowup as a driving mechanism of turbulence in shell models.
\newblock {\em Phys. Rev. E}, 87(5):053011, 2013.

\bibitem{mailybaev2014continuous}
A.A. Mailybaev.
\newblock Continuous representation for shell models of turbulence.
\newblock {\em arXiv:1409.4682}, 2014.

\bibitem{obukhov1941spectral}
A.~M. Obukhov.
\newblock {Spectral energy distribution in a turbulent flow}.
\newblock {\em Dokl. Akad. Nauk SSSR}, 32(1):22--24, 1941.

\bibitem{ohkitani1989temporal}
K.~Ohkitani and M.~Yamada.
\newblock {Temporal intermittency in the energy cascade process and local
  Lyapunov analysis in fully developed model of turbulence}.
\newblock {\em Prog. Theor. Phys.}, 89:329--341, 1989.

\end{thebibliography}
 
\end{document}